\documentstyle[aps]{revtex}
\begin{document}
\twocolumn[
{\bf U. Leonhardt and P. Piwnicki reply to the
``Comment on `Relativistic Effects of Light in Moving Media with
Extremely Low Group Velocity ' '' by M. Visser}\\
\smallskip
]

We are grateful to Matt Visser for clarifying the interpretation
of optical black holes \cite{LPliten}. Waves in moving media may
become trapped if the flow outruns the wave velocity, an effect
which may establish artificial black holes or equivalent sonic
analogs in superfluids and alkali Bose-Einstein condensates
\cite{Sound}. However, as Visser's Comment \cite{Comment} points
out clearly, the medium should flow towards a drain, in order to
form a black hole. The flow should guide light into the drain
such that it disappears beyond an event horizon, provided of
course that the flow velocity is sufficient for a horizon to form.

In our Letter \cite{LPliten} we had chosen a radially spinning
vortex as our theoretical model for a suitable flow, because a
vortex allows to combine two intriguing aspects of slow light
\cite{Hau} in moving media \cite{Slow}, the optical
Aharonov--Bohm effect and the analog of gravitational attraction.

The Aharonov--Bohm effect of slow light in Bose--Einstein
condensates may facilitate, for the first time, the observation
of the long--range nature of quantum vortices. Previously, only
vortex cores have been seen directly, by the trapping of
electrons at vortex lines in superfluid $^4{\rm He}$ \cite{Tilley}
or by taking pictures of expanded droplets of alkali
Bose--Einstein condensates carrying vortices \cite{Madison}. In
superfluid $^3{\rm He}$, the texture of vortex matter has been
inferred using NMR \cite{Tilley}.

Slow light in moving quantum fluids experiences phase shifts due
to the Doppler detuning of an atomic resonance, with ultrahigh
motion sensitivity \cite{Gyro}. Phase-contrast microscopy of
Bose--Einstein condensates \cite{Andrews} can be applied to
measure local phase shifts and to retrieve the flow pattern from
the phase profile, as a form of optical tomography. Consequently,
apart from the exciting prospects of forming artificial black
holes, slow light can serve as a major experimental tool to
explore in situ quantum fluids.

\bigskip
\noindent
U. Leonhardt\\
School of Physics and Astronomy\\
University of St Andrews\\
North Haugh\\
St Andrews, Fife, KY16 9SS, Scotland\\

\noindent
P. Piwnicki\\
Physics Department\\
Royal Institute of Technology (KTH)\\
Lindstedtsv\"agen 24\\
S-10044 Stockholm, Sweden\\



\end{document}